\documentclass{PoS}
\usepackage{epsfig}
\def\gsi{\raise0.3ex\hbox{$>$\kern-0.75em\raise-1.1ex\hbox{$\sim$}}}
\newcommand{\gsim}{\mathop{\gsi}}

\PoS{PoS(LAT2005)283}

\title{Testing Topology Conserving Gauge Actions \\for Lattice QCD}

\ShortTitle{Testing Topological Conserving Gauge Action for Lattice QCD}

\author{\speaker{Kei-ichi ~Nagai}, \, Karl Jansen\\
John von Neumann-Institute f\"ur Computing NIC,\\
DESY, Platanenallee 6, D-15738 Zeuthen, Germany\\
Email: \email{Keiichi.Nagai@desy.de, Karl.Jansen@desy.de}
}

\author{
Wolfgang ~Bietenholz, \, Luigi ~Scorzato \\
Institut f\"ur Physik, Humboldt Universit\"{a}t zu Berlin \\ 
Newtonstr. 15, D-12489 Berlin, Germany \\
Email: \email{bietenho@physik.hu-berlin.de, Luigi.Scorzato@physik.hu-berlin.de}}

\author{
Silvia ~Necco \\
Centre de Physique Th\'{e}orique, Luminy, Case 907 \\ 
F-13288, Marseille Cedex 9, France \\
Email: \email{necco@cpt.univ-mrs.fr}}

\author{
Stanislav ~Shcheredin \\
Fakult\"{a}t f\"{u}r Physik, Universit\"{a}t Bielefeld \\
D-33615 Bielefeld, Germany \\
Email: \email{shchered@physik.hu-berlin.de}}

\abstract{
We explore gauge actions for lattice QCD,
which are constructed such that the occurrence of
small plaquette values is strongly suppressed.
Such actions originate from the admissibility condition 
in order to conserve the topological charge.
The suppression of small plaquette values is expected
to be advantageous for numerical studies in the $\epsilon$-regime
and also for simulations with dynamical quarks.
Performing simulations at a lattice spacing of about 0.1 fm,
we present numerical results for the static potential,
the physical scale $r_0$,
the stability of the topological charge history,
the condition number of the kernel of the overlap operator
and the acceptance rate against the step size in the local HMC algorithm.
}

\FullConference{XXIIIrd International Symposium on Lattice Field Theory\\
25-30 July 2005\\
Trinity College, Dublin, Ireland}

\begin{document}
\section{Introduction and motivation}
\label{sec:intro}
\vspace*{-2mm}
{\em Chiral perturbation theory} ($\chi$PT)~\cite{XPT}
and lattice QCD are powerful tools
to extract quantities at low energy which are relevant for QCD.
Lattice QCD simulations can provide the determination of 
the {\em Low Energy Constants (LECs)} of $\chi$PT from first principles 
calculations. 
A particular situation is found
when one enters the $\epsilon$-regime~\cite{eps-reg}, where
\begin{equation}
m_\pi \sim \frac{1}{L^2} 
\quad, \quad 
m_{\pi}^{-1} > L \quad , \quad \left(L \gg \frac{1}{2 F_\pi} \right) \, .
\end{equation}
There are analytical formulae from $\chi$PT 
that describe the behavior of physical quantities in the 
$\epsilon$-regime as a function of the volume and the quark mass. 
These formulae are parameterized by the {\em infinite volume}
{\em LECs} of the effective chiral Lagrangian. 
Therefore we can extract physically relevant information 
even from the unphysical $\epsilon$-regime.
As a peculiarity of the $\epsilon$-regime, 
since one is working in small boxes~($L \gsim 1.1~{\rm fm} \cdots 1.5~{\rm fm}$), 
observables depend significantly on the topological sector,
and predictions exist for expectation values in specific sectors~\cite{LeuSmi}.
For the parameters that have been used in the $\epsilon$-regime simulations,
it would be of particular interest 
to collect large sets of configurations 
\footnote{The topologically neutral sector is problematic due to the frequent
appearance of very small Dirac eigenvalues, which leads to strong
spikes in the Monte Carlo histories of correlation functions \cite{AA}.}
with an index $\vert \nu \vert >0$.

In general, however,
it is not obvious to define topological sectors on the lattice.
A neat definition exists with overlap fermions satisfying 
the Ginsparg--Wilson relation~\cite{GW,Neu1}
\begin{equation}
D_{\rm ov}^{(0)}   \gamma_{5} +  \gamma_{5} D_{\rm ov}^{(0)} = 
\frac{1}{\mu} D_{\rm ov}^{(0)} \gamma_{5} D_{\rm ov}^{(0)} \ , \quad 
D_{\rm ov}^{(0)} =\mu \Big[ 1 + \gamma_{5} Q / \sqrt{Q^{2}} \Big] 
 \ , \quad
Q = \gamma_{5} (D_{\rm W} - \mu) \ ,\label{overlap}
\end{equation}
where $\mu \gsim 1$ is a mass parameter in the operator $Q$.
Since $D_{\rm ov}^{(0)}$ has exact zero modes 
with definite chiralities~\cite{Neu1,Has,ML},
one can use the index as a definition of the topological charge
due to the Atiyah--Singer theorem.

While overlap fermions define the fermion sector of lattice QCD, 
if one insists on exact lattice chiral symmetry, 
the lattice gauge action of QCD can be constructed 
in many ways~\cite{gauge}.
One requires 
the naive continuum limit for all lattice gauge actions to coincide, 
in which case they fall into the same universality class. 
The simplest formulation is the {\em Wilson plaquette action}
\begin{equation}  \label{Wilact}
S_{\rm W}[U] = \beta \sum_{P} S_{P}(U_{P}) \ , \quad
S_{P}(U_{P}) = 1 - \frac{1}{3} {\rm Re~Tr}  U_{P} \ ,
\end{equation}
where the sum over $P$ runs over all plaquettes.

For the overlap operator (\ref{overlap}),
the topological transitions are excluded under continuous deformations
if all the plaquette variables $U_{P}$ in the configurations
involved obey the inequality~\cite{luschact,HJL}
$S_{P}(U_{P}) < \varepsilon = \frac{2}{5d(d-1)} = \frac{1}{30} \,
\, ({\rm for}\ {\rm all}\ P)$.
Later on H.\ Neuberger showed a more tolerant bound~\cite{Neu2}
$\varepsilon = \frac{1}{(1 + 1/\sqrt{2})d(d-1)} \simeq \frac{1}{20.5}$.
For this {\em admissibility condition}, 
the exponential locality of the Ginsparg--Wilson fermions 
is guaranteed~\cite{HJL},
and the topological charge is conserved rigorously.
However, it is very difficult to realize this condition 
in a numerical simulation. For practical purposes
we have to relax $\varepsilon$ to larger values 
than the analytical bounds.

\section{Proposal of the gauge actions}
\label{sec:gauge}
\vspace*{-2mm}
We now describe a number of non-standard lattice gauge actions,
which suppress the unwanted small plaquette values leading 
to transitions between different topological sectors.
The naive continuum limit of these actions
coincides with that of Wilson action~(\ref{Wilact}). 
Based on the proposal by M.~L\"uscher~\cite{luschact},
we consider topology conserving gauge actions~\cite{prelim}
\begin{eqnarray}  
&&S_{\varepsilon ,n}^{\rm hyp} (U_{P}) = 
\frac{S_{P}(U_{P})}{ [ 1 -  S_{P}(U_{P}) / \varepsilon ]^{n}} \quad
\,\,\, {\rm for} \,\,\,  S_{P}(U_{P}) < \varepsilon \,\,\,  \quad
{\rm and} \quad + \infty  \,\,\,\, {\rm otherwise}  \,,
\label{hypact} \, \\
&&S_{\varepsilon ,n}^{\rm pow} (U_{P}) = S_{P}(U_{P})
+ \frac{1}{\varepsilon} S_{P}(U_{P})^{n} 
\label{powact} \ , \\
&&S_{\varepsilon ,n}^{\rm exp} (U_{P}) = S_{P}(U_{P}) \cdot
\exp \{ S_{P}(U_{P})^{n} / \varepsilon \} \, ,  \qquad \qquad \qquad (n>0) \,.
\label{expact}
\end{eqnarray}

In simulations, 
the action~(\ref{hypact}) with $n=1$ was first used 
in the Schwinger model by Fukaya and Onogi \cite{FuOn}. 
They set $\varepsilon =1$,
i.e.\ far above the theoretical value of about $0.29$, 
but they still observed topological stability over hundreds of configurations.
Here we investigate this type and its extension 
to the gauge action of lattice QCD.

\begin{figure}[htbp]
\vspace*{-3mm}
\begin{minipage}{0.5\hsize}
\begin{center}
\epsfig{angle=0,file=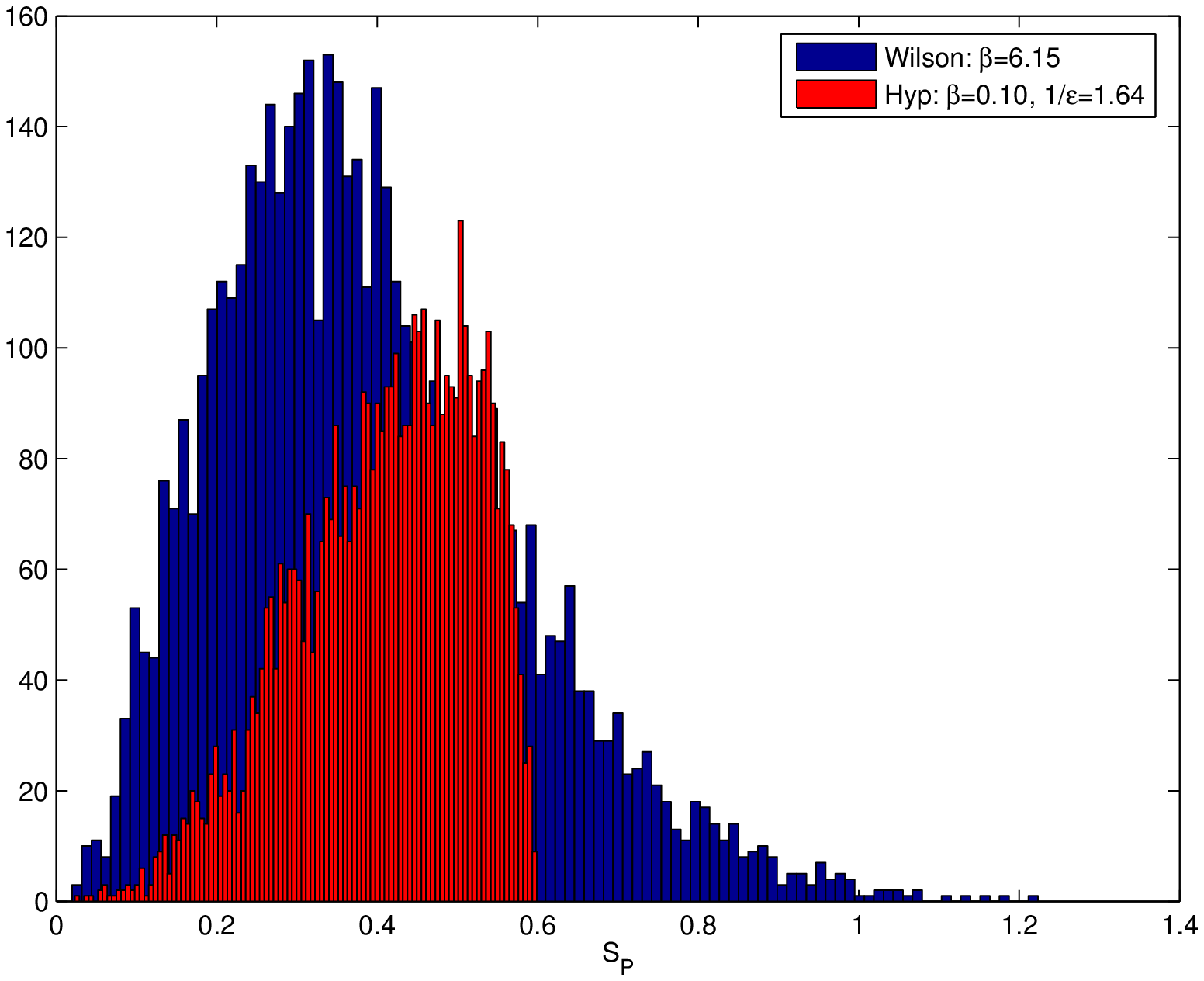, width=\textwidth}
\end{center}
\end{minipage}
\begin{minipage}{0.5\hsize}
\begin{center}
\epsfig{angle=-90,file=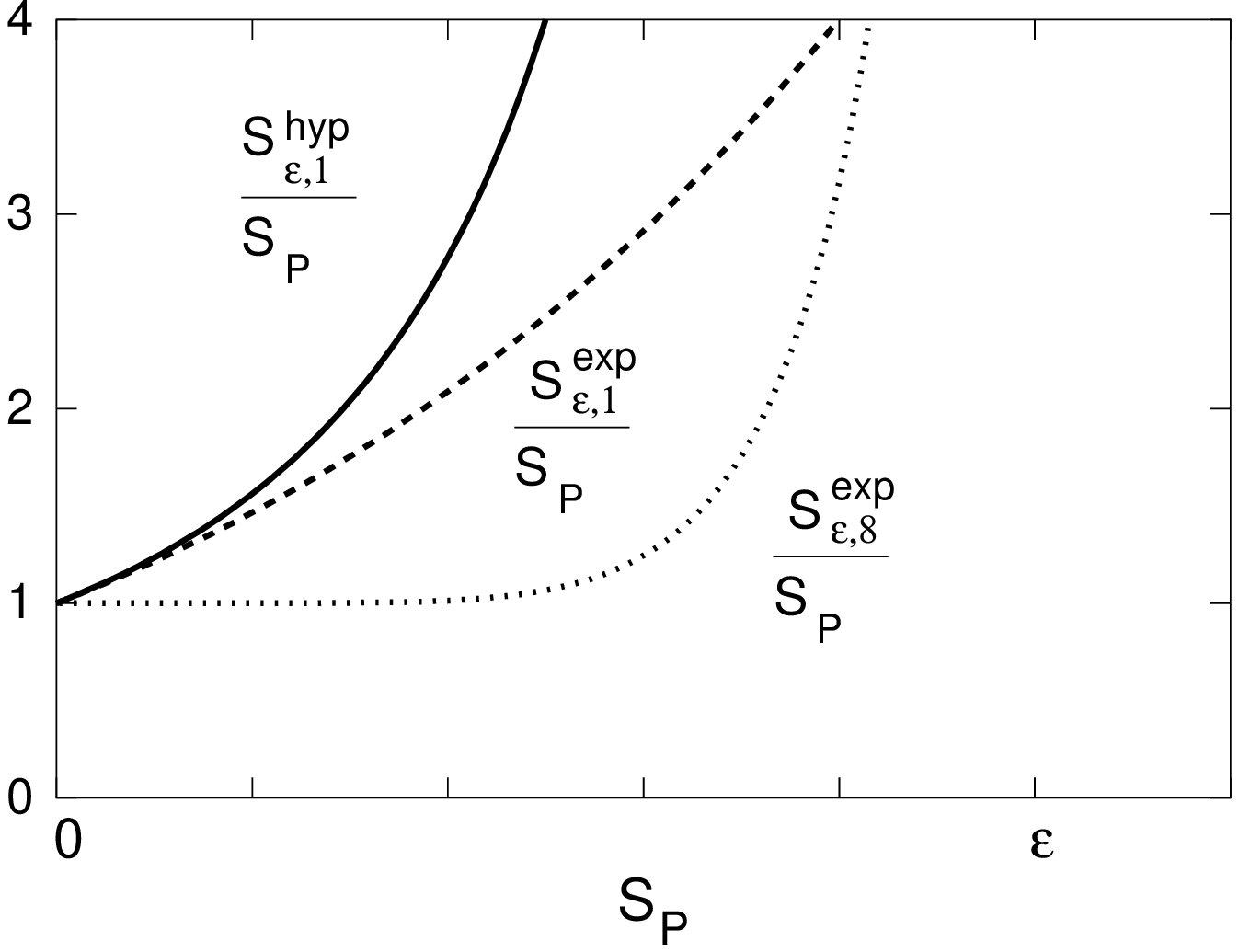, width=0.85\textwidth}
\end{center}
\end{minipage}
\caption{
Left: Histograms of $S_P$ for the topology conserving gauge action,
compared to the Wilson action, on a $4^4$ lattice.
Right: Ratio of the HMC force
between the Wilson action $S_P$
and modified actions.
}
\label{fig:action}
\end{figure}

We show the basic properties of these actions.
The left plot of fig.~\ref{fig:action} 
is the histogram of plaquette values 
for the action $S_{\varepsilon ,n=1}^{\rm hyp}$,
compared to the Wilson action, on a $4^4$ lattice.
In the topology conserving gauge action,
the occurrence of very small plaquette values is drastically suppressed.
The right plot of fig.~\ref{fig:action}
shows the ratio of the force in the HMC algorithm 
between the Wilson plaquette action $S_P$
and topology conserving gauge actions.
The action $S_{\varepsilon ,n}^{\rm exp}$ makes 
a sharp wall {\em continuous},
but shows the same behavior 
as the Wilson action over a wide range.

For the generation of the configurations we used 
a {\em local HMC} algorithm.
Since $S_{\varepsilon ,n}^{\rm hyp}$ is 
non-linear for the link variables,
the heatbath and over-relaxation algorithms are not 
straightforwardly applicable.
We use a $16^4$ lattice
and measure physical quantities every 50 trajectories.
See table~\ref{tab:hyp} about the acceptance rate
of local HMC. At $1/\varepsilon=1.64$,
the acceptance rate is about $65\%$ 
at the Molecular dynamics step $d \tau=0.1$.
However the acceptance improves as $d \tau$ decreases.

\section{Monte Carlo history of $Q_{top}$: Stability of the topological charge.}
\label{sec:qtopo}
\vspace*{-2mm}
\begin{figure}
\hspace*{-0.3cm}
\epsfig{angle=0,file=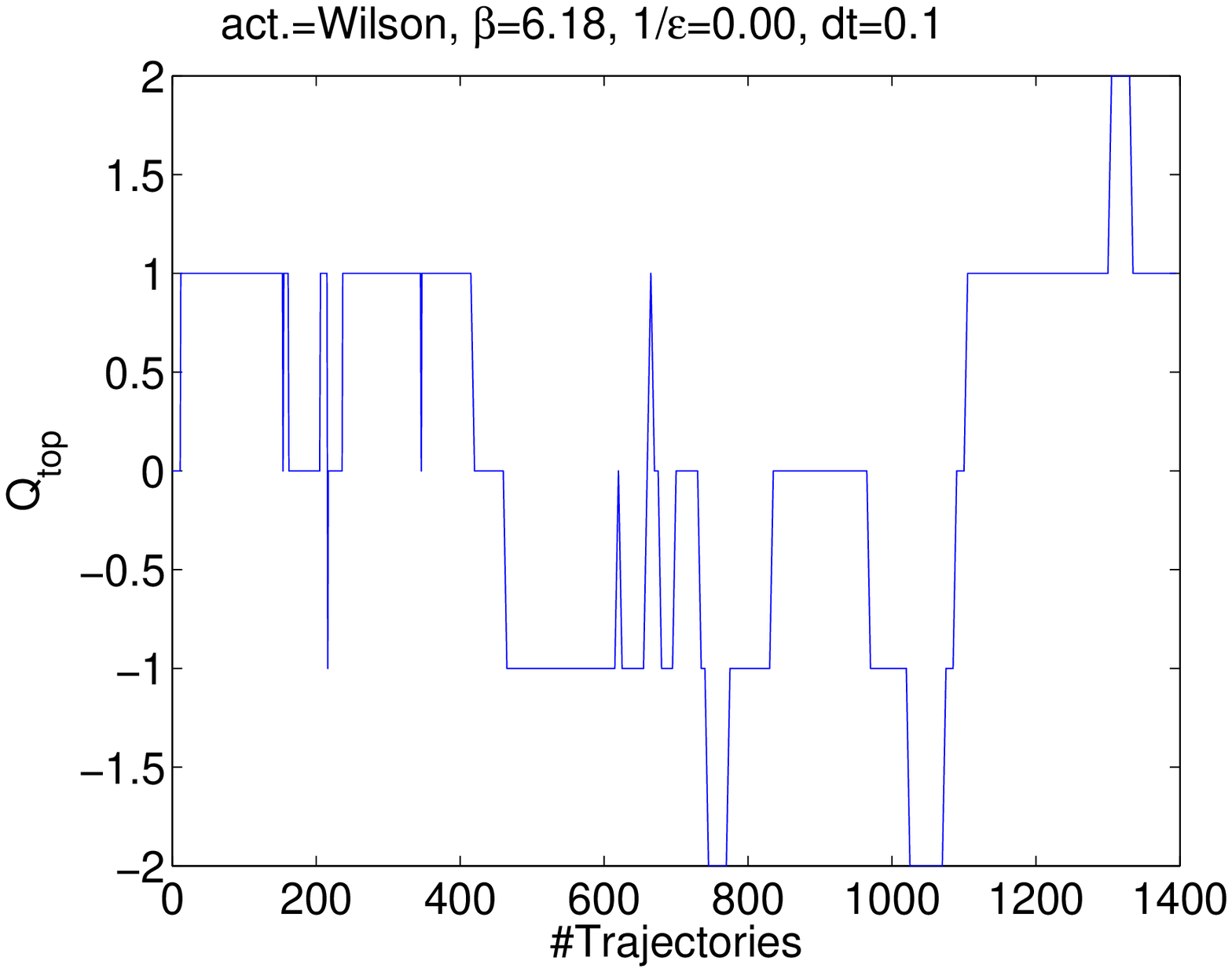, width=.53\textwidth}
\hspace{-0.6cm}
\epsfig{angle=0,file=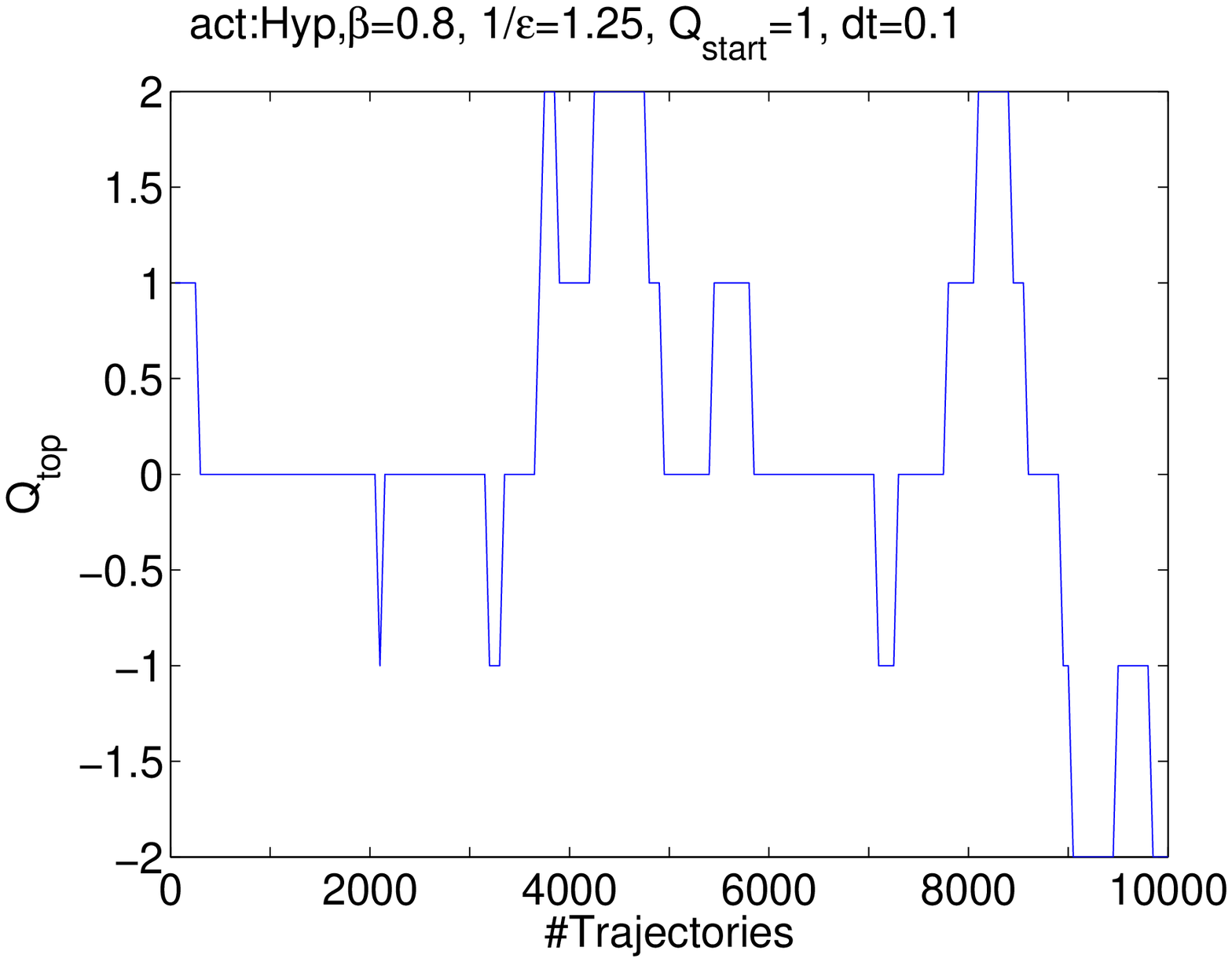, width=.53\textwidth}
\caption{MC history of $Q_{top}$ for 
$S_{\rm W}$ (left) and $S^{\rm hyp}_{\epsilon =0.8,n=1}$ (right).}
\label{fig:tophist}
\end{figure}
In fig.~\ref{fig:tophist}
we show the Monte Carlo history of the topological charge $Q_{top}$,
which is evaluated by the cooling method\footnote{The cooling charges
agree in practically all cases with the overlap indices
in the range $\mu=1.3, ..., 1.6$, 
as we verified for a subset of the configurations.},
for the Wilson action and a modified gauge action.
The topological charge in $1/\varepsilon \ll 20.5$
is not always conserved.
However, the changes of $Q_{top}$ are suppressed.
Here we define the quantity representing 
the stability of the topological charge: 
$f_{top}=$(the number of jumps of $Q_{top}$)$/$(the number of trajectories),
see also table~\ref{tab:hyp}.
The value $f_{top}$ for the modified gauge actions
is 10 times smaller than for the Wilson action
and thus these actions clearly 
stabilize the topology charge.
Also the autocorrelation of the plaquette $\tau^{plaq}$
is getting shorter as $1/\varepsilon$ increases.

\section{The static potential, the physical scale and lattice artifacts}
\label{sec:statpot}
\vspace*{-2mm}
In this section we evaluate the physical scale and the lattice artifacts
for various parameters of $\beta$ and $1/\varepsilon$
in the modified gauge actions.
The most established method of setting a scale in pure gauge theory
is the measurement of the static potential and the force
at relatively large distances, extracted from Wilson loops,
see {\em e.g.} Ref.~~\cite{SNe}.

\begin{figure}
\hspace{-0.5cm}
\epsfig{angle=270,file=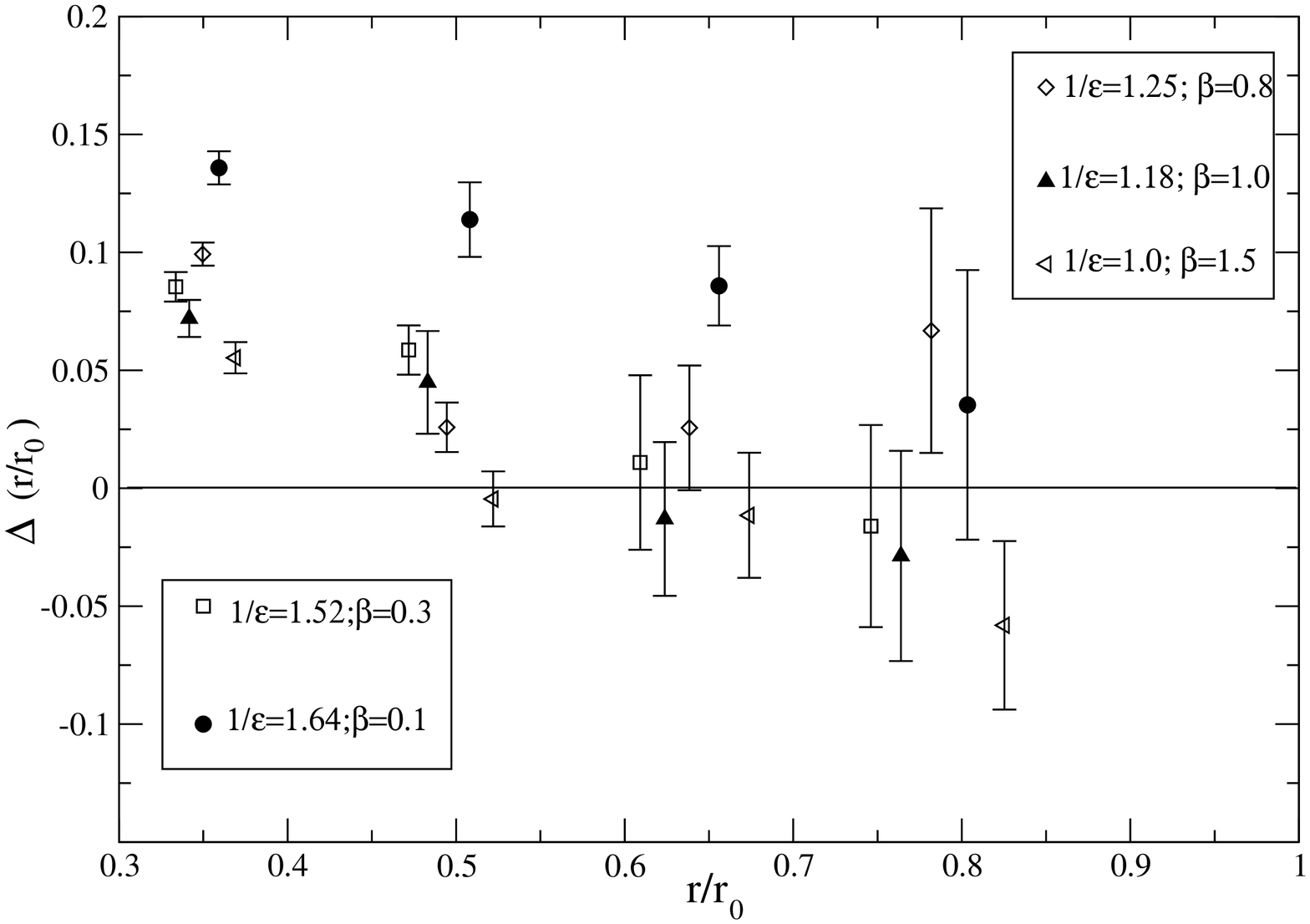, width=.55\textwidth}
\hspace{-0.7cm}
\epsfig{angle=270,file=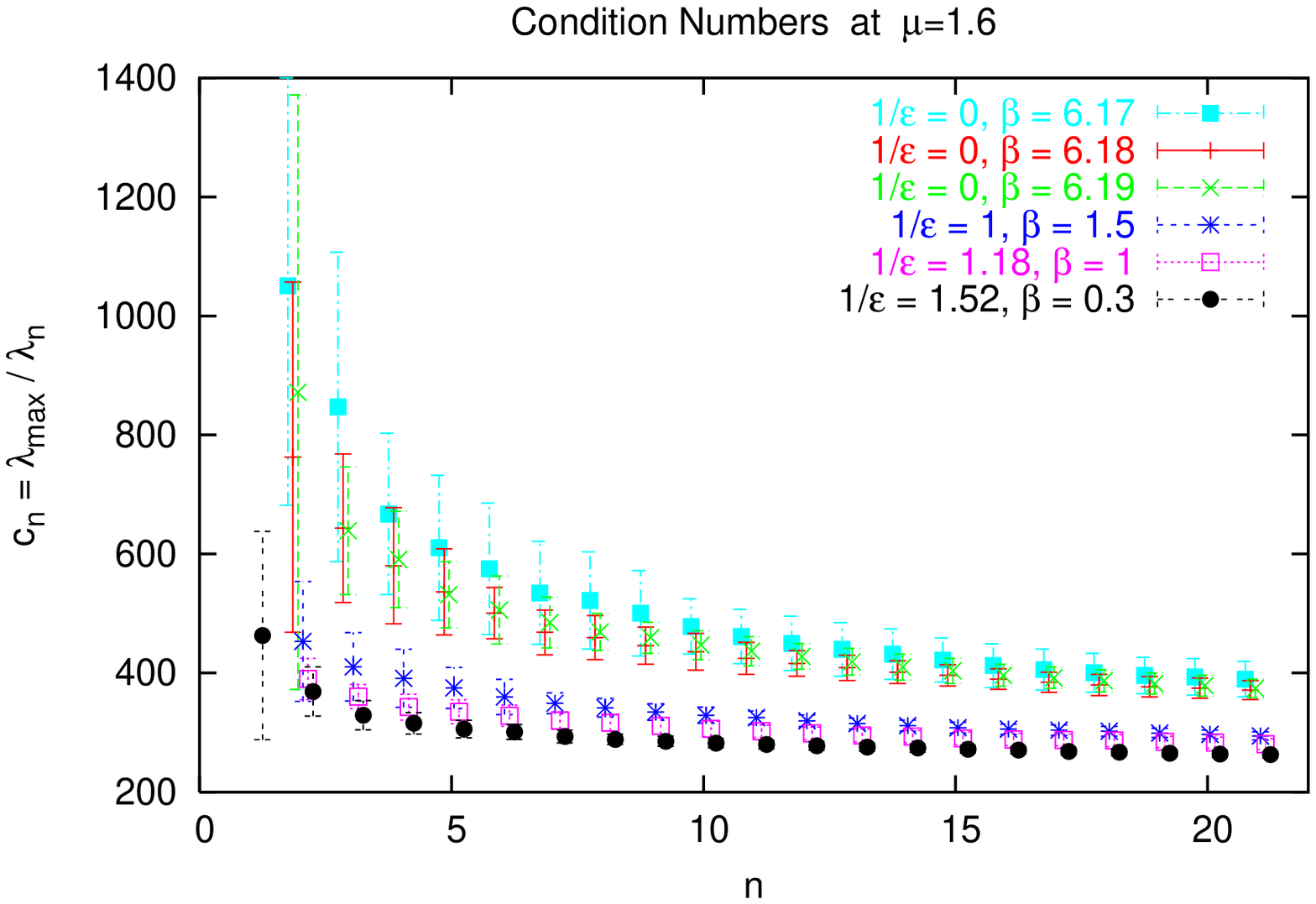, width=.55\textwidth}
\caption{Left: Lattice artifact by the standard estimate
for the action $S_{\varepsilon ,n=1}^{\rm hyp}$.
Right: The overlap condition numbers $c_{2} ... c_{21}$ at $\mu=1.6$.}
\label{fig:scale1}
\end{figure}

Results are shown in the left plot of fig.~\ref{fig:scale1}.
A way to check the lattice artifacts~\cite{SNe}
is to compare the short distance force 
at finite lattice spacing 
with the one extrapolated to the continuum limit $r^2 F(r/r_{0})|_c$
\begin{equation}  \label{ratforce}
\Delta(r/r_0)=\frac{r^2 F(r/r_0)-r^2 F(r/r_0)|_c}{r^2 F(r/r_0)|_c} \ .
\end{equation}
As seen in this figure,
for increasing $1/\varepsilon$,
the discretization errors increase as well.
Typically the artifacts of $S^{\rm hyp}_{\varepsilon,n=1}$ 
at $1/\varepsilon \gsim 1$ are about $10\%$.
Even in $1/\varepsilon=1.64$,
the lattice artifact is less than $15\%$.
This is comparable with the Iwasaki and DBW2 gauge actions.
Note that the static potential at distance $r \sim r_0$
has not been determined with high precision
(this is reflected in large uncertainties 
on the quantities $\Delta(r/r_0)$).


\section{Condition number for the overlap operator}
\label{sec:condnumber}
\vspace*{-2mm}
We also evaluate the effect of topology conserving gauge actions
on the condition number of the overlap operator~(\ref{overlap}).
We observe the condition number defined as
$C_n=\frac{\lambda_{max}}{\lambda_n}$
where $\lambda_{max}$ is the maximum eigenvalue and 
$\lambda_n$ is the $n$'th eigenvalue of the kernel $Q^2$ 
in the overlap operator.
$C_n$ is relevant 
if $n-1$ modes of $Q^2$ are projected out.
Fig.~\ref{fig:scale1} (right plot)  shows 
the behavior
of the condition numbers $c_{2}, \cdots ,  c_{21}$ at $\mu=1.6$
and about $r_0/a \sim 7.0$.
We see an improvement which increases 
if only a few modes are projected out 
and if $1/\varepsilon$ grows.
Hence the computation of $Q/\sqrt{Q^2}$ 
in the overlap operator gets fast
by using topology conserving gauge actions.

\section{Summary}
\label{sec:summary}
\vspace*{-2mm}
Simulations in the $\epsilon$-regime of chiral perturbation theory
using the overlap operator can be simplified when appropriate 
choices of the gauge actions are made. 
We tested a number of gauge actions that suppress small plaquette values
and investigated the properties of these actions~\cite{article}.
Our results can be summarized as follows, see also 
table~\ref{tab:hyp}. 

\begin{itemize}
\item The actions of eqs.~(\ref{hypact},~\ref{powact},~\ref{expact})
stabilize the topological charge
and could therefore be profitable in QCD simulations
in the $\epsilon$-regime.

\item The actions of eqs.~(\ref{powact},~\ref{expact})
are conceptually clean, have a positive
transfer matrix 
and show lattice artifacts that are acceptable and
compare well with those of the Iwasaki and DBW2 gauge actions. 

\item The condition number of the kernel of the overlap operator 
is improved
by using the topology conserving gauge actions 
when compared to the standard Wilson plaquette action. 
This speeds up the simulations of overlap fermions in the quenched 
approximation and will also help in 
dynamical simulations using the (global) HMC algorithm.


\end{itemize}


\vspace*{-2mm}
\noindent
{\small
{\bf Acknowledgements} \  We thank H.\ Fukaya, M. L\"{u}scher 
and T.\ Onogi for useful communications. 
This work was supported by the Deutsche Forschungsgemeinschaft 
through SFB/TR9-03. Part of
the computations were performed on the IBM p690 clusters of the HLRN
``Norddeutscher Verbund f\"ur Hoch- und H\"ochstleistungsrechnen'' (HLRN) 
and at NIC, Forschungszentrum J\"{u}lich.}

\begin{table}
\begin{center}
\begin{tabular}{|c|c|c|c|c|c|c|c|}
\hline
$\varepsilon^{-1}$  & $\beta$ & $r_0/a$ & $\beta_W$ & $d \tau$ 
& $\tau^{\rm plaq}$ & $f_{\rm top}$ & acc.\ rate \\
\hline
\hline
0. & 6.19 & 7.14(3) & 6.19  & 0.1    & 7(1)   & 2.2e-2 & $>$ 99 \% \\
\hline
1. & 1.5 & 6.6(2) & 6.13(2) & 0.1   & 2.2(1) & 2.4e-3 & $>$ 99 \% \\
1. & 1.5 & 6.6(2) & 6.13(2) & 0.01  & 2.2(1) & 3.2e-3 & $>$ 99 \% \\
\hline
1.18 & 1. & 7.2(2) & 6.18(2) & 0.1  & 1.2(1) & 1.6e-3 & $>$ 99 \% \\
1.18 & 1. & 7.2(2) & 6.18(2) & 0.02/0.01 & 1.3(1) & 1.4e-3 & $>$ 99 \% \\
\hline
1.25 & 0.8 & 7.0(1) & 6.17(1) & 0.1 & 1.1(1) & 2.5e-3 & $>$ 99 \% \\
\hline
1.52 & 0.3 & 7.3(4) & 6.19(4) & 0.1 & 0.8(1) & 9.4e-4 & $\approx$ 95 \% \\
\hline
1.64 & 0.1 & 6.8(3) & 6.15(3) & 0.1 & 1.0(1) & 7.0e-4 & $\approx$ 65 \% \\
1.64 & 0.1 &        &         & 0.05& 0.7(1) & 2.3e-3 & $\approx$ 78 \% \\
1.64 & 0.1 &        &         &0.025& 0.6(1) & 3.5e-3 & $\approx$ 93 \% \\
\hline
\end{tabular}
\end{center}
\caption{Results for $S^{\rm hyp}_{\varepsilon,1}$
at various values of $\varepsilon$ and $\beta$, on a $16^{4}$ lattice. 
$\beta_W$ is the $\beta$ value
of Wilson action corresponding to the measured ratio $r_0/a$.
$\tau^{plaq}$ is the autocorrelation time of the plaquette value.}
\label{tab:hyp}
\end{table}

%
%
%
%

\end{document}